\documentclass[preprint,12pt]{elsarticle}
\usepackage{mathrsfs}

\usepackage{lineno}

\usepackage{amsfonts} 
\usepackage{xcolor}

\usepackage{threeparttable}

\usepackage[english]{babel} 

\usepackage{multicol} 

\usepackage{amsmath} 
\usepackage{mathtools} 
\usepackage{amssymb} 

\usepackage{amsthm} 

\usepackage[T1]{fontenc} 
\usepackage{textcomp} 
\usepackage[utf8]{inputenc} 

\usepackage[colorlinks=true,citecolor=blue,linkcolor=black,urlcolor=blue]{hyperref} 

\usepackage[left=1.5cm,top=1.5cm,right=1.5cm,bottom=1.5cm]{geometry} 

\usepackage{txfonts} 
\usepackage{enumerate} 

\usepackage{float} 

\usepackage{graphicx} 
\usepackage{caption} 
\usepackage{subcaption} 
\usepackage{siunitx}

\usepackage{booktabs}

\usepackage{lipsum}   

\newtheoremstyle{teoremas}
{\topsep}                    
{\topsep}                    
{\itshape}                   
{}                           
{\bfseries}                  
{}                          
{ }                       
{}  
\theoremstyle{teoremas}

\newtheoremstyle{definiciones}
{\topsep}                    
{\topsep}                    
{}                   				
{}                           
{\bfseries}                   
{}                          
{ }                       
{}  
\theoremstyle{definiciones}

\newtheoremstyle{obsyejemp}
{\topsep}                    
{\topsep}                    
{}                   				
{}                           
{\scshape}                   
{:}                          
{ }                   
{}  													

\theoremstyle{obsyejemp}

\newtheorem{thm}{Theorem}

\newdefinition{rmk}{Remark}
\newproof{pf}{Proof}

\begin{document}

\begin{frontmatter}



\title{Effects of a Differentiating Therapy on Cancer-Stem-Cell-Driven Tumors}


\author[inst1,inst2]{J. Fotinós}

\address[inst1]{Instituto de Física Enrique Gaviola, CONICET, Córdoba, 5000, Córdoba, Argentina}

\author[inst1,inst2]{L. Barberis}
\author[inst1,inst2]{C. A. Condat}

\address[inst2]{Universidad Nacional de Córdoba (FaMAF, UNC), Bvd. Medina Allende s/n, Ciudad Universitaria, Córdoba, 5000, Córdoba, Argentina}

\begin{abstract}
The growth of many solid tumors has been found to be driven by chemo- and radiotherapy-resistant cancer stem cells (CSCs). A suitable therapeutic avenue in these cases may involve the use of a differentiating agent (DA) to force the differentiation of the CSCs and of conventional therapies to eliminate the remaining differentiated cancer cells (DCCs). To describe the effects of a DA that reprograms CSCs into DCCs, we adapt a differential equation model developed to investigate tumorspheres, which are assumed to consist of jointly evolving CSC and DCC populations. We analyze the mathematical properties of the model, finding the equilibria and their stability. We also present numerical solutions and phase diagrams to describe the system evolution and the therapy effects, denoting the DA strength by a parameter \(a_{dif}\).To obtain realistic predictions, we choose the other model parameters to be those determined previously from fits to various experimental datasets. These datasets characterize the progression of the tumor under various culture conditions. Typically, for small values of \(a_{dif}\) the tumor evolves towards a final state that contains a CSC fraction, but a strong therapy leads to the suppression of this phenotype. Nonetheless, different external conditions lead to very diverse behaviors. For some environmental conditions, the model predicts a threshold not only in the therapy strength, but also in its starting time, an early beginning being potentially crucial. In summary, our model shows how the effects of a DA depend critically not only on the dosage and timing of the drug application, but also on the tumor nature and its environment.
\end{abstract}


\begin{highlights}
\item The action of a differentiating therapy on stem cell cultures is modeled.
\item A strong differentiating agent can suppress the cancer stem cell phenotype.
\item The outcome of the differentiation therapy depends critically on the tumor
environment.
\item The importance of the starting time of the therapy is assessed.
\item The potentialities and limitations of the differentiation therapy are exhibited.
\end{highlights}


\end{frontmatter}


\section{Introduction}
		The cancer stem cell hypothesis states that cancer growth is driven by a subpopulation of CSCs that have the ability to self-renew and differentiate, giving rise to the DCCs that comprise the tumor bulk \cite{Baccelli2012, Batlle2017,Jagust2019}. They can also reversibly enter quiescent states and resist radiotherapy and cytotoxic drugs, which helps to explain tumor recurrence and metastasis \cite{Dean2005, Lacerda2010, Ogawa2013, Kreso2014, Batlle2017, Prieto-Vila2017, LaPorta2018, Malta2018, Feng2020, Bhattacharya2020}. The hypothesis has been backed up by the identification of CSCs in a growing and diverse group of tumors. This has motivated the search for new therapeutic paradigms based on the idea that the incapacitation of the CSCs, with the simultaneous use of conventional therapies to reduce the DCC load, is the most effective procedure to control tumor growth \cite{Lacerda2010, Li2017, Jagust2019, ALHulais2019, Taniguchi2020}. One possible therapeutic course to eliminate the CSC component is to induce CSC differentiation. Retinoic acids (all-trans retinoic acid - ATRA, 9-cis retinoic acid, and 13-cis retinoic acid) have shown potential as differentiating agents. Unfortunately, in the case of solid tumors these promising results have not yet been transformed into effective therapies \cite{Jin2017, Costantini2020, Giuli2020, Hen2020, Prasad2020, Jin2022}. One of the reasons for this failure is our deficient understanding of the processes involving cancer stem cells and their reaction to external interventions. 

        The growing recognition of the importance of understanding the processes underpinning CSC-fueled tumor growth has led to the formulation of a number of mathematical models \cite{Agur2010,Ganguly2006, Michor2008, Sole2008, Turner2009, Rodriguez-Brenes2011, LaPorta2012, Gao2013, DosSantos2013, Liu2013, Hannezo2014, Zhou2014, Enderling2015, Weiss2017, Forster2017, Alqudah2020, Meacci2021, Barberis2021a, Fischer2022,Swanson2022}. These models offer insights into growth and differentiation rates, cell population fractions, lateral inhibition, and chemo- and radio-therapy effects, to cite a few processes. Simultaneously, the complexity of the cancer phenomenon has also led to the development of simplified biological models to investigate diverse tumor properties under better-defined experimental conditions. Tumorspheres, spheroids grown from single-cell suspensions obtained from permanent cell lines or tumor tissue, are particularly useful to investigate CSC–driven tumor growth. They allow us to investigate the role of CSCs in tumor growth without the interference of complicating factors \cite{Reynolds1996, Gu2011, Chiodi2011, Yang2013, Weiswald2015, Lee2016}. Mathematical models can also help us to extract valuable information from tumorsphere experiments. However, the usual spheroid models, such as that in Ref. \cite{Delsanto2005} are not particularly well-suited for the task. For this reason, we have developed a two-population model for tumorsphere growth \cite{Benitez2019, Barberis2021}. This model exhibits a transcritical bifurcation, where a purely non-stem-cell attractor is replaced by a new attractor that contains both CSCs and DCCs. It allows us to reconstruct the time evolution of the CSC fraction, which is usually not directly measured. Application of this model to the experiments of Chen and coworkers on tumorspheres formed out of three cancer lines \cite{Chen2016} showed that, while intraspecific interactions are usually inhibitory, interspecific interactions stimulate growth \cite{Benitez2019}. Later, the model was used to interpret the results of experiments performed in Tianjin under different mechanical and growth factor conditions \cite{Wang2016}. These confirm that niche memory is responsible for the characteristic population dynamic observed in tumorspheres \cite{Benitez2021}.  
        
        The goal of this paper is to obtain information about the efficacy of differentiation therapy when applied to the simplest nontrivial system involving CSCs, i.e., the tumorsphere. In this pursuit, we extend the model of \cite{Benitez2021} to include the effects of a DA using as model parameters those obtained from the application of the model to experiments performed in the absence of the agent. Since we will not include any other therapies in our discussion, we will consider the differentiation therapy successful if it forces the tumorsphere to evolve towards a pure DCC system. 
        
        In the next section, we present the model and describe its mathematical properties. In Section \ref{sec: numerical predictions for experimental cases} we apply it to predict the modifications induced by the DA on the evolution of tumorspheres grown under well-defined culture conditions, paying special attention to the roles of dosage and timing. This will be done using values for the model parameters in the absence of therapy obtained from fits to experimental datasets \cite{Barberis2021,Benitez2019,Benitez2021}.
		
		\section{The model and its properties}
            \subsection{The model}
		We will describe the evolution of a tumorsphere subject to differentiation therapy using a system of two coupled ordinary differential equations. The model is an extension of the one developed by Benítez et al. \cite{Benitez2019,Benitez2021} to study tumor progression in the absence of therapy. The equations describe the evolution of the total number of CSCs and DCCs, which will be denoted by \(S\) and \(D\), respectively.

  		\begin{figure}[h]
			\centering
			\includegraphics[width=0.7\textwidth]{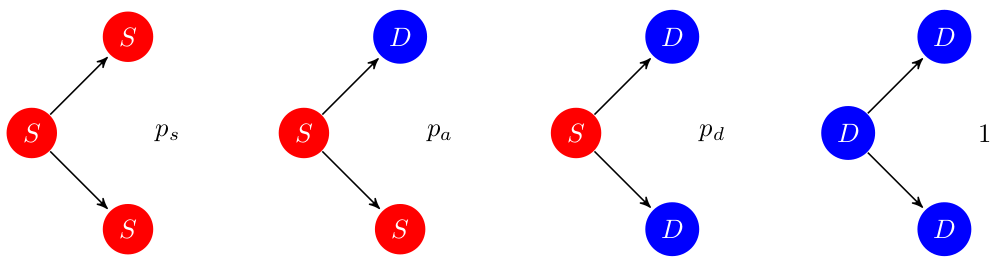}
			\captionof{figure}{\emph{Cell reproduction routes.} CSCs are shown in red, with the possible results of their reproduction. From left to right, a \(S\) cell can generate two \(S\) cells with probability \(p_s\), one \(S\) cell and one \(D\) cell (in blue) with probability \(p_a\) or two \(D\) cells with probability \(p_d\). DCCs can only self-replicate with probability 1.} \label{figure: posibilidades reproduccion}	
		\end{figure}

        Whereas \(D\) cells can only self-replicate, \(S\) cells display three options: they can self-renew, yielding two \(S\) cells, with a probability \(p_s\); they can yield two DCCs with a probability \(p_d\); and they can reproduce asymmetrically, yielding one CSC and one DCC, with probability \(p_a = 1 - p_s - p_d\) (see Fig. \ref{figure: posibilidades reproduccion}).
            
		In general, one might consider that the CSC and DCC subpopulations reproduce at different rates. Nevertheless, here we consider a unique reproduction rate \(r\) for both populations. This is a natural simplification given that, in the experiments, it is not possible to discriminate between the individual rates. Furthermore, the consideration of separate growth rates would not change the position of the equilibria. This is a well-known feature of the continuous models for population dynamics: unless the environment is patchy, the role of the interactions between the populations is dominant, and the growth rate simply defines the temporal scale of the system (\cite{Britton2003}).

\begin{figure}[h]
        \centering
        \includegraphics[width=0.5\textwidth]{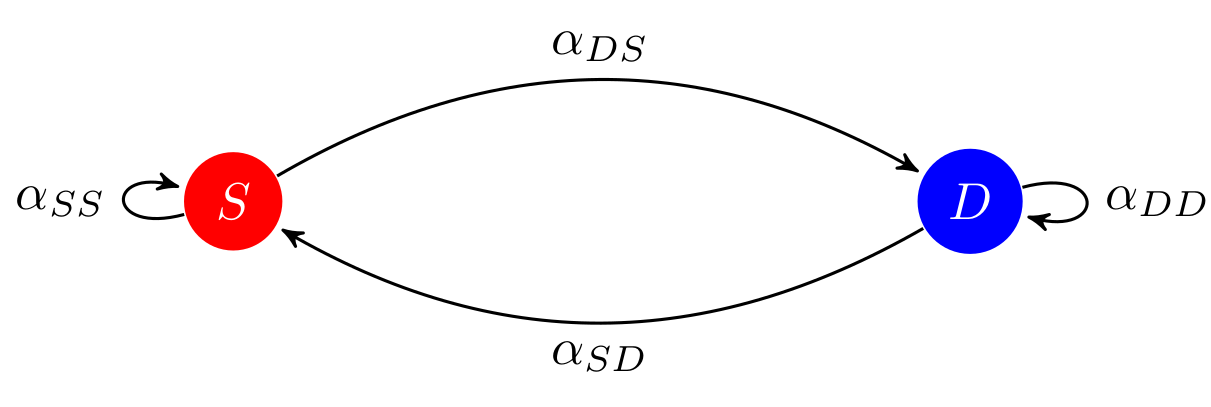}
        \captionof{figure}{\emph{Interaction coefficients.} Each coefficient is placed next to the arrow indicating the direction of the corresponding interaction. Thus, for example, \(\alpha_{SS}\) represents the intraspecific interaction between \(S\) cells. Similarly, \(\alpha_{DS}\) describes the interspecific action of \(S\) cells on \(D\) cells.} \label{figure: esquema interaccion especies}	
    \end{figure}

		Since the behavior of a given cell is influenced by the rest of the cells of its kind (intraspecific interactions), and by cells of the other subpopulation (interspecific interactions), the model has the typical interacting species structure \cite{Britton2003}. These interactions are represented by coefficients \(\alpha_{ij}\) that account for the effect of subpopulation  \(j\) over subpopulation \(i\). Thus, intraspecific interactions are represented by coefficients \(\alpha_{ij}\) with \(i=j\), while interspecific interactions are described by coefficients with \(i \neq j\). This is illustrated in Fig. \ref{figure: esquema interaccion especies}.

		If \(\alpha_{ij}>0\) supopulation \(j\) inhibits the growth of subpopulation \(i\). This inhibition may be due to competition for space, nutrients, or other resources. If, on the contrary, \(\alpha_{ij}<0\), subpopulation \(j\) stimulates the growth of subpopulation \(i\). This collaboration may be due, for instance, to attempts to restore a putative state of equilibrium.

		The equations of the model, where \(\Pi=p_d/p_s\) and the dot indicates the temporal derivative, are as follows,
		
		\begin{subequations}
			\label{eq: tumesf terap}
			\begin{align}
				\dot{S} &= r p_s S (1 - \Pi - \alpha_{SS} S - \alpha_{SD} D) - a_{dif} S \label{eq: S tumesf terap} \\
				\dot{D} &= [r D +r S (p_a + 2 p_d)] (1 - \alpha_{DS} S - \alpha_{DD} D ) + a_{dif} S \label{eq: D tumesf terap}
			\end{align}
		\end{subequations}
		
		In Eq. \eqref{eq: S tumesf terap} for the number of \(S\) cells, the first term contains the factor \(r p_s S\) that, multiplied by \(1-\Pi=(p_s-p_d)/p_s\), can be thought of as the effective intrinsic growth rate of the \(S\) subpopulation. Reproductions that yield two \(S\) cells increase the \(S\) population, and reproductions that result in two \(D\) cells reduce that population. It should be noted that asymmetric reproductions do not change the number of \(S\) cells. The other terms in the parenthesis are the interaction terms. As stated above, the term involving \(\alpha_{SS}\) takes care of the intraspecific interaction between \(S\) cells, while the term including \(\alpha_{SD}\) describes the interspecific action of \(D\) cells over the \(S\) cells.
		
		Switching now our attention to equation \eqref{eq: D tumesf terap} for the growth of \(D\) cells, we see that the bracket corresponds to the reproduction of these cells, considering the contribution of the \(D\) cells themselves, and the contribution of the \(S\) cells, whose reproduction paths are weighted accordingly. As in the previous equation, this factor multiplies the one representing the interactions.
		
		The last term in both equations describes the therapy, whose effect is to differentiate a fraction \(a_{dif}\) of the \(S\) cells, turning them into \(D\) cells. By adding both equations, one can easily see that therapy does not change the total number of cells. It just transfers cells from \(S\) to \(D\). Of course, the therapy term is nonzero only at the times the DA is present in the system.

		It is often convenient to use a non-dimensional version of the equations. This allows us to reduce the number of parameters and is usually convenient when searching for numerical solutions. A non-dimensional version of Eqs. \eqref{eq: tumesf terap} is given by
		
		\begin{subequations}
			\label{eq: tumesf terap ADIM}
			\begin{align}
				\dot{X} &= X (P-AX-BY)-M(1-P)X \label{eq: X tumesf terap}\\
				\dot{Y} &= (X+Y)(1-CX-Y)+MX \label{eq: Y tumesf terap}
			\end{align}
		\end{subequations}
		where,
		\begin{align*}
			&X = \alpha_{DD} (p_a + 2 p_d) S  &Y = \alpha_{DD} D  \\
			&A = \frac{\alpha_{SS}}{\alpha_{DD}} \frac{p_s}{p_a + 2 p_d}  &B = \frac{\alpha_{SD}}{\alpha_{DD}} p_s \\
			&C = \frac{\alpha_{DS}}{\alpha_{DD}} \frac{1}{p_a + 2 p_d}  &P = p_s - p_d  \\
			&M = \frac{a_{dif}}{r(p_a+2p_d)}  &\tau = r t 
		\end{align*}
		The dot now stands for the derivative with respect to \(\tau\).
		

		\subsection{General properties}
		Next, we will discuss the general properties of the model. The critical points of the dynamical system \eqref{eq: tumesf terap}, i.e., its equilibrium states, will be described as points in the phase space \((S,D)\). There are four of them, the origin \(T_0=(0, 0)\), which corresponds to the total absence of cells; the \(D\)-cell equilibrium \(T_1=(0, 1/\alpha_{DD})\), where there are no \(S\) cells and the number of \(D\) cells equals its environmental carrying capacity; and the two equilibrium states \(T_{2,3}=(S_{2,3}, D_{2,3})\), where both populations coexist.
		
		A stability analysis shows that the eigenvalues of the linearisation matrix about \(T_0\) are
		\begin{align*}
			\lambda_1^{(0)} &= r \\	
			\lambda_2^{(0)} &= r (p_S - p_D) - a_{dif}
		\end{align*}
		Since \(r>0\), this point is always unstable. It may be a repulsor (negative attractor) or a saddle point. This depends on whether the effective reproduction of the \(S\) cells exceeds the differentiation induced by the therapy. It is interesting to note that, if \(T_0\) is originally a repulsor, we can turn it into a saddle point by increasing the therapy efficiency \(a_{dif}\).
		
		A similar analysis for \(T_1\) shows that the eigenvalues are
		\begin{align*}
			\lambda_1^{(1)} &= - r \\	
			\lambda_2^{(1)} &= r p_s \left(1 - \Pi - \frac{\alpha_{SD}}{\alpha_{DD}} \right) - a_{dif}
		\end{align*}
		
		Moreover, the eigenvector corresponding to \(\lambda_1^{(1)}\) has the direction of the \(D\) axis, showing that this equilibrium will always be stable along this axis. Thus, \(T_1\) will be an attractor or a saddle point, depending on the values of the parameters. More concretely, in the absence of therapy, for \(T_1\) to be an attractor the \(D\) cells must exert a stronger inhibitory action on the \(S\) cells than over the \(D\) cells. If it is initially a saddle point, the application of the therapy allows to turn \(T_1\) into an attractor by increasing the therapy parameter \(a_{dif}\). This is important because it tells us that, with a high enough dosage, we can force the system to evolve towards a state without CSCs. The resulting tumor could be dealt with with classical therapies, for it lacks the resistance granted by the CSC subpopulation.
		
		Both the explicit expressions for the coexistence states and those for their eigenvalues and eigenvectors are too cumbersome to shed any light on the positivity or stability of these equilibria. We will pay special attention to these states later, when some sets of experimental values for the growth parameters are introduced. Next, we prove instead two theorems for the model. The first one deals with the positivity of the solutions: reasonable \(S(t)\) and \(D(t)\) must remain in the first quadrant of the phase space \(\forall t \). The second yields several properties of the solutions. In particular, it gives a condition for the initial growth of the CSC subpopulation.

        \begin{thm} \label{Teo: extension teo 1}
        Let \(S(t)\) and \(D(t)\) be a pair of solutions to the system  \eqref{eq: tumesf terap} for the initial conditions \(S(0)=S_0>0\) and \(D(0)=D_0=0\). Then \(S(t)\) is non-negative \(\forall t>0\). If, additionally, we assume that \(p_s < 1 \lor a_{dif} \neq 0\), then \(\alpha_{DS}<\xi(S_{MAX})\) is enough to ensure the non-negativity of \(D(t)\), where \(S_{MAX}\) is the maximum of \(S(t)\) and
		\begin{align*}
			\xi(S)=\frac{1}{S}\left[1+ \frac{a_{dif}}{r(p_a+2p_d)}\right] 
		\end{align*}
        \end{thm}

        \begin{pf}
        We start with the positivity of \(S\). Let us assume that \(S(t)\) becomes non-positive for the first time at \(t_1>0\), and negative after that. Then \(S(t_1)=0 \land \dot{S}(t_1)<0\), but, from Eq. \eqref{eq: S tumesf terap}, we have that \(S(t_1)=0 \implies \dot{S}(t_1)=0\), which is a contradiction that comes from the assumption that \(\exists ~ t_1\) with the demanded properties. Therefore, under the theorem assumptions, \(S(t)\) cannot be negative.
			
        For the positivity of the \(D\) subpopulation, let us notice that \( (p_s \neq 1 \lor a_{dif} \neq 0) \land \alpha_{DS} <\xi(S_{MAX})\) is a sufficient condition to ensure initial growth. At early times, we can neglect terms involving a factor \(D\) in Eq. \eqref{eq: D tumesf terap} and  write
        \begin{align*}
            \dot{D} \approx r S (p_a + 2 p_d) (1 - \alpha_{DS} S) + a_{dif} S 
        \end{align*}
        This can be bounded from below by replacing \(\alpha_{DS}\) by \(\xi(S_{MAX})\) (because we showed that \(S>0\) and by hypothesis \(\alpha_{DS}<\xi(S_{MAX})\)). We thus get
        \begin{align*}
            \dot{D} > S [r (p_a + 2 p_d) + a_{dif}] \left( 1 - \frac{S}{S_{MAX}} \right)
        \end{align*}
        which is always greater than zero, and therefore the initial growth of the \(D\) population is proven.
        It follows that there is at least an initial interval where \(D(t)>0\). Let us assume that \(t_2\) is the smallest positive time such that \(D=0\), and after that,  \(D\) becomes negative. Therefore, \(\dot{D}(t_2)<0\). Using Eq. \eqref{eq: D tumesf terap},
        \begin{align*}
            \dot{D}(t_2) &=[r D(t_2) + \\ &+ r S (p_a + 2 p_d)] (1 - \alpha_{DS} S - \alpha_{DD} D(t_2) ) + a_{dif} S \\
            &= [r (p_a + 2 p_d)(1 - \alpha_{DS} S ) + a_{dif}]  S < 0
        \end{align*}
        Since \(S>0\), this implies
        \begin{align*}
            r (p_a + 2 p_d)(1 - \alpha_{DS} S ) < - a_{dif} \\ \implies \alpha_{DS} > \xi(S) > \xi(S_{MAX})
        \end{align*}
        which is an absurd. This follows from the assumption that \(\exists t_2\) such that \(D\) loses its non-negativity. Therefore, \(D(t) \geq 0, \forall t>0\).
        \end{pf}

            \begin{rmk}
            The positivity of \(D\) is assured if \(\alpha_{DS}<0\), for \(\xi(S_{MAX})>0, \forall S_{MAX}>0\). Such condition is likely to be fulfilled because we expect the CSCs to promote the growth of the DCCs. This is a sufficient but not necessary condition, unless the CSC population can grow without limit, because in that case \(\xi(S_{MAX}) \longrightarrow 0\). This tells us that if the S population remains bounded, we can allow some inhibition of the D cells by their CSC counterparts.
            \end{rmk}

            \begin{rmk}
                The positivity condition is more easily met for greater \(a_{dif}\). The reason for this is that \(\xi(S)\) grows with \(a_{dif}\), and therefore there are more values of \(\alpha_{DS}\) that satisfy \(\alpha_{DS}<\xi(S_{MAX})\).
            \end{rmk}
		
		\begin{thm} \label{Teo: extension teo 2}
			Suppose the evolution of a system is described by the equations \eqref{eq: tumesf terap ADIM}, with initial conditions \(X(\tau=0)=X_0\) e \(Y(\tau=0)=Y_0\). Then,\\
			a)  If the initial conditions (seed) is small enough, such that \(|AX_0|,|CX_0|,|BY_0|,|Y_0| \ll 1 \land P \neq 1\), then initially (short-time behavior)
					$$ M<\frac{P}{1-P} \implies ~~ \frac{\partial |X|}{\partial \tau} >0 ~ , ~~ \frac{\partial |Y|}{\partial \tau} >0$$
					$$ M>\frac{P}{1-P} \implies ~~ \frac{\partial |X|}{\partial \tau} <0 ~ , ~~ \frac{\partial |Y|}{\partial \tau} >0 $$
			b) \(X_0 = 0 \implies X(\tau)=0, \forall \tau>0 \) \\
			c) \(X_0 = 0, Y_0 > 0 \implies \lim_{\tau \to \infty} Y(\tau)=1\)
		\end{thm}
	
		\begin{pf} $\\ $
		    a) Under these hypotheses, we may use a linearized version of the equations. The resulting equation for \(X\) can be solved analytically, giving
				$$ X(\tau) = X_0 ~ \text{exp}{\left\lbrace [P-M(1-P)]\tau\right\rbrace} $$
				Since, by definition, \(P \in [-1, 1]\) and, by hypothesis, \(P \neq 1\), the argument of the exponential is positive if \(M<P/(1-P)\).
				Since \(\text{sgn}(X(\tau))=\text{sgn}(X_0)\), if the argument of the exponential is positive, \(\partial_\tau |X| > 0\).
				
				Solving the linearized equation for \(Y\), we get
				\begin{align*}
					Y(\tau)= e^\tau \left[ Y_0 + \frac{X_0}{1-P} \left( 1 - e^{(M+1)(1-P)\tau} \right)  \right]
				\end{align*}
				Substituting to obtain \(Y(X)\) and deriving with respect to \(X\), we obtain an equation for the trajectory in the \((X,Y)\) plane:
				\begin{align*}
					\frac{\partial Y}{\partial X} = \frac{\left[ X(\tau)/X_0\right]^{\frac{M+1}{\frac{P}{1-P}-M}}}{1-(M+1)(1-P)} \left( \frac{Y_0}{X_0}  + \frac{1}{1-P} \right)  - \frac{1}{1-P}
				\end{align*}
				Defining 
				\begin{align*}
					\beta = \left[ \frac{X(\tau)}{X_0} \right]^{\frac{M+1}{\frac{P}{1-P}-M}} \text{ and } \gamma = 1-(M+1)(1-P) \text{,}
				\end{align*}
				we have, by definition, \(\gamma<1\). If, additionally
				\begin{align*}
					\gamma>0 \implies M<\frac{P}{1-P} \implies \frac{X(\tau)}{X_0}>1 \\ \implies \beta>1 \implies \frac{\beta}{\gamma}>1
				\end{align*}
				Since \(\text{sgn}(Y_0)=\text{sgn}(X_0)\) and
				\begin{align*}
					\frac{\partial Y}{\partial X} = \frac{\beta}{\gamma} \frac{Y_0}{X_0} + \frac{1}{1-P} \left(\frac{\beta}{\gamma} - 1 \right) 
				\end{align*}
				we can finally assert that
				\begin{align*}
					M<\frac{P}{1-P} \implies ~~ \frac{\partial Y}{\partial X} >0
				\end{align*}
				On the other hand, if
				\begin{align*}
					M>\frac{P}{1-P} \implies \gamma<0 \implies ~~ \frac{\partial Y}{\partial X} <0
				\end{align*}
				We can link the sign of the temporal derivative of \(Y\) to the sign of the temporal derivative of \(X\). This and the assumption that \(M < P/(1-P)\) lead to the inequality \(\partial_\tau |X| > 0\), which completes the proof. Moreover, it is immediate to find the explicit initial form of the trajectory in the \((X,Y)\) plane:
                \begin{align*}
                    Y(\tau)= \left[ Y_0 + \frac{X_0}{1-P} \right] \left[ \frac{X(\tau)}{X_0} \right]^{\frac{1}{P-M(1-P)}} - \frac{X_0}{1-P} \left[ \frac{X(\tau)}{X_0} \right]
                \end{align*}
                b) This follows from the fact that the origin of the phase space is an equilibrium state of the system.\\
		      c) If \(X_0=0\), using the statement proven above, the equation for \(Y\) is \(\dot{Y} = Y (1-Y)\). For positive values of \(Y\), there is a unique stable equilibrium: \(Y=1\).
		\end{pf}
		
		\begin{rmk}
		    This theorem is also valid if \(\alpha_{DD}<0\), that is, when the system grows in the third quadrant of the non-dimensional phase space.
		\end{rmk}

            \begin{rmk}
                In terms of the original parameters, the threshold value of the therapy parameter for initial growth is \(a_{dif} = r(p_s-p_d)\). In the absence of therapy, \(p_s<p_d\) is sufficient for the CSCs not to thrive (independently of the value of \(r\)). Under treatment, it will be necessary that \(p_s>p_d+a_{dif}/r\) for the \(S\) population to grow. Note that the suppressor effect of  \(a_{dif}\) is stronger when \(r\) is small. Furthermore, if \(a_{dif}/r>1\), there is no \(p_s\) that can satisfy the condition for initial growth. This result will be adequate for short-times and small seeds, when interactions are truly negligible. Perhaps this might not seem very interesting, since one would expect the therapy to start some time after the onset of growth, but there are at least two cases where it would be relevant. The first case would occur if therapy is applied during the generation of a secondary tumor (metastasis), possibly started by a CSC invading healthy tissue. The second case would take place when, after a conventional therapy, only a few CSCs survive. In both cases, growth could be easily contained if \(a_{dif}>r\).
            \end{rmk}

            \begin{rmk}
                The initial condition \((S_0,D_0)=(1,0)\) does not necessarily constitute a small seed. This is because the non-dimensional seed size depends on the carrying capacity of the DCCs. Therefore, if \(K_D = 1/\alpha_{DD}\) is big compared to the initial population, the seed will be small.
            \end{rmk}

	\section{Numerical predictions for experimental cases} \label{sec: numerical predictions for experimental cases}
	   As mentioned before \cite{Barberis2021,Benitez2019,Benitez2021}, the model in the absence of therapy was used to fit data from experiments in which tumorspheres were cultured under different environmental conditions. Here, we use Eqs. \eqref{eq: tumesf terap} with the parameters obtained from these fits to analyze the system dynamics in terms of the therapy strength \(a_{dif}\). The cases we considered correspond to the experiments carried out by Chen et al. \cite{Chen2016}, and Wang et al. \cite{Wang2016}.

\begin{table}
  \centering
  \begin{tabular}{@{} l *{4}{S[table-format=2.4]} @{}}
\toprule
{\text{Constant}} & {\text{Chen}} & {\text{Hard (Wang)}} & {\text{Soft (Wang)}} \\
        \midrule
        {\(r \text{ [1/days]}\)}   & 1.32 & 0.1335 & 0.0685\\
        {\(p_s\)}  & 0.36 & 0.7124 & 0.9701\\
        {\(p_d\)}  & 0.160 & 0.0000 & 0.0019\\
        {\(\alpha_{SS}\)}  & 0.0519 & -0.0456 & 0.0873\\
        {\(\alpha_{SD}\)}  & -0.032816 & -0.5280 & -0.4185\\
        {\(\alpha_{DS}\)}  & -0.0175 & -0.1376 & -0.2061\\
        {\(\alpha_{DD}\)}  & 0.020616 & 1.8329 & 0.3668\\
\bottomrule
\end{tabular}
\caption{Fitted values of the parameters for the three experimental cases}
\label{tab: param}
\end{table}

	\subsection{Experiments of Chen et al.} \label{Subsection Chen}
		In these experiments \cite{Chen2016}, a microfluidic platform was used to grow single-cell derived spheres. This platform consisted of a chip formed by a set of microchambers with a non-adherent surface coating. This array provided robust single-cell isolation and allowed tumorspheres to grow freely in the absence of external tensions. The authors reported the size of the spheres during the first 10 days of growth. Here we will focus on those grown out of the T47D line. Barberis et al. \cite{Barberis2021} fitted the corresponding data, obtaining the values of the model parameters shown in Table \ref{tab: param}. As we can see from the signs of the interaction parameters, intraspecific interactions (\(i=j\)) turned out competitive, whereas interspecific interactions (\(i \neq j\)) were cooperative. Using the parameters in Table \ref{tab: param}, we can now analyze tumorsphere dynamics as a function of therapy efficiency.
		
\begin{figure}
        	\centering
        	\begin{subfigure}[b]{0.49\textwidth}
        		\centering
        		\includegraphics[width=1\textwidth]{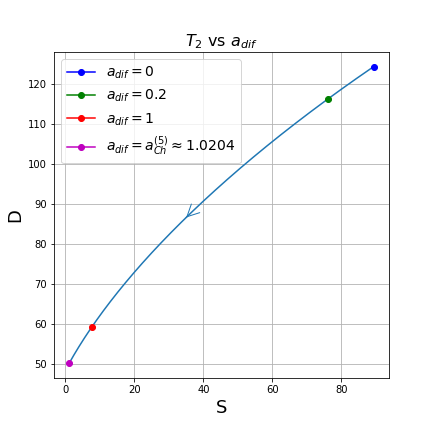}
        		\captionof{figure}{\emph{Coexistence \(T_2(a_{dif})\) in case 1.}} \label{figure: trayectoria T2 vs adif, Chen}
        	\end{subfigure}
        	\hfill
        	\begin{subfigure}[b]{0.49\textwidth}
        		\centering
        		\includegraphics[width=1\textwidth]{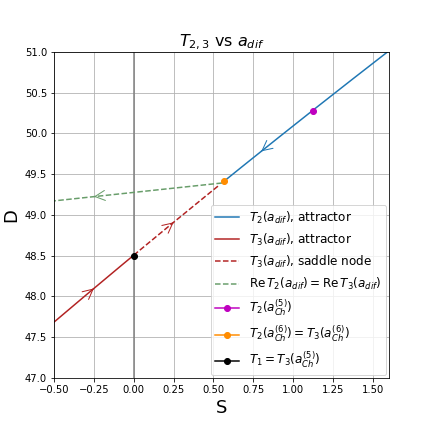}
        		\captionof{figure}{\emph{Coexistence \(T_{2,3}(a_{dif})\) in case 2.}} \label{figure: trayectoria T23 vs adif, Chen}
        	\end{subfigure}
        	\caption{\emph{Position of the equilibria in terms of \(a_{dif}\) [Chen].} The arrows indicate the displacement direction of the equilibria when increasing \(a_{dif}\). Panel \ref{figure: trayectoria T2 vs adif, Chen} shows the curve \(T_2(a_{dif})\) in case 1, for which every initial condition leads to this coexistence state. The size of the resulting tumorsphere and the CSC fraction decrease with increasing \(a_{dif}\). Panel \ref{figure: trayectoria T23 vs adif, Chen} shows the states \(T_{2,3}\) in: the end of case 1-B, where \(T_2\) and \(T_3\) are attractors in the first and second quadrants, respectively, passing through case 2-A, where \(T_3\) becomes a saddle node in the first quadrant, and ending in case 2-B. This case begins when \(T_2\) meets \(T_3\) at \(a_{dif}=a^{(6)}_{Ch}\), and continues for higher values of \(a_{dif}\). There, \(T_2\) and \(T_3\) become complex conjugates whose real parts are shown as a green dashed line.}
        	\label{fig: puntos críticos chen}
            \end{figure}

		The possible behaviors of the tumorsphere under a differentiating therapy are summarized in Table \ref{tab: regimenes chen}. We find that \(T_0\) can be a repulsor or a saddle point, depending on whether \(a_{dif}\) is respectively greater or smaller than \(a_{Ch}^{(1)}\simeq 0.2640\): in any case, the tumor cannot be completly eliminated, irrespective of therapy strength. We also find that \(T_1 \simeq (0,48.5)\) is a saddle node if \(a_{dif}<a^{(5)}_{Ch}\simeq1.0204\); if \(a_{dif}\) is above this threshold, \(T_1\) becomes an attractor, and the CSCs are completely eliminated. The coexistence states \(T_{2,3}\) are real only for \(a_{dif}<a_{Ch}^{(6)}\simeq 1.0206\). For stronger therapies, they become complex conjugates and play no physical role. When \(T_2\) is real, it is also an attractor in the first quadrant. \(T_3\) starts from the fourth quadrant, at \(a_{dif}=0\); then at \(a_{Ch}^{(1)}\), it enters the second quadrant through the origin \(T_0\); at \(a_{Ch}^{(5)}\simeq 1.0204\) it passes into first quadrant, crossing \(T_1\), the equilibrium of \(D\) cells. Both bifurcations, at \(a_{Ch}^{(1)}\) and \(a_{Ch}^{(5)}\), are transcritical. This implies that the critical points meeting there exchange their stability. The evolution of the equilibria locations under changes in \(a_{dif}\) is shown in Fig. \ref{fig: puntos críticos chen}.
		
	\begin{table*}
    	\centering
	\begin{tabular}{ |p{1.5cm}||p{2cm}|p{2.5cm}|p{2.5cm}|p{2.5cm}|p{2.5cm}|  }
		\hline
		Case& Range & \(T_0\) &\(T_1\)&\(T_2\)&\(T_3\)\\
		\hline
		1-A & \([0,a^{(1)}_{Ch})\)  & Repulsor & Saddle Node & Attractor & Non-biological\\
		1-B & \((a^{(1)}_{Ch},a^{(5)}_{Ch})\) & Saddle Node & Saddle Node & Attractor & Non-biological\\
		2-A & \((a^{(5)}_{Ch},a^{(6)}_{Ch})\) & Saddle Node & Attractor & Attractor & Saddle Node\\
		2-B & \((a^{(6)}_{Ch},\infty)\) & Saddle Node & Attractor & Non-biological&  Non-biological\\
		\hline
	\end{tabular}
	\caption{Summary of the qualitative behavior of a T47D tumorsphere for the various ranges of \(a_{dif}\), with the parameters of the experiment of Chen et al. \cite{Chen2016}.}
	\label{tab: regimenes chen}
    \end{table*}

        It is useful to think that there are two general cases, with two sub-cases each, as indicated in Table \ref{tab: regimenes chen}. In case 1 (\(a_{dif}<a^{(5)_{Ch}}\)), any initial condition leads to a coexistence state that gets smaller as the therapy parameter \(a_{dif}\) increases. This coexistence begins at \(T_2(a_{dif}=0)\simeq (89.3, 124.3)\), and stops being real at \(T_2(a_{dif}=a^{(6)}_{Ch})\simeq (0.57, 49.4)\). This can be seen in Fig. \ref{figure: trayectoria T2 vs adif, Chen}. The only difference between sub-cases A y B of case 1, is the change in stability of the origin, due to the transcritical bifurcation that takes place when \(T_3\) and \(T_0\) meet for \(a_{dif}=a^{(1)}_{Ch}\). This has no significant impact on the system dynamics.

        Broadly speaking, one might say that in case 2 \(a_{dif}\) has increased so that the stable equilibrium is composed only by DCCs. A more detailed analysis reveals that this is not entirely the case. As a matter of fact, in sub-case A both \(T_1\) and \(T_2\) are attractors, whose basins of attraction are divided by the stable manifold of \(T_3\). Nonetheless, given the very narrow domain of \(a_{dif}\) [\(a_{dif} \in (1.0204, 1.0206)\)] in subcase 2-A and the proximity between the stable equilibria, the simplification made above ends up being quite reasonable. In subcase 2-B, the coexistence states disappear and all trajectories starting from initial conditions in the first quadrant converge to \(T_1\).

        This behavior is to be expected since intraspecific competition and interspecific cooperation usually give rise to a stable coexistence. If the therapy strength increases, the number of CSCs of the stable equilibrium decreases, and so does their promotion of the DCCs, resulting in a decline of both sub-populations. When \(a_{dif}\) exceeds the threshold \(a_{Ch}^{(6)}\), the differentiation is such that the populations cannot coexist in a stable state, and \(T_1\) becomes the only stable equilibrium. It is interesting to note that there are less DCCs in \(T_1\) than in \(T_2\) for any \(a_{dif}<a_{Ch}^{(6)}\). The cooperation of the CSCs allows the DCCs to reach a population greater than their intrinsic environmental carrying capacity.

        The effect of the therapy is to reduce the number of CSCs, eliminating them completely if the therapy is strong enough. The starting time of the therapy does not play a relevant role in this case; all that seems to matter is the dosage or therapy efficiency \(a_{dif}\). Also note that growth is always bounded. We take this case as a benchmark, mainly for three reasons:
        \begin{enumerate}
            \item The tumorspheres grow freely, in the absence of external tensions.
            \item There are no stem cell growth factors added to the culture medium.
            \item As explained in \cite{Chen2016}, the experimental array used minimizes typical problems of tumorsphere cultures.
        \end{enumerate}
        
        Comparison of this case with others enable us to link changes in culture conditions to modifications in the resulting dynamics.

        \begin{figure}[h]
    	\centering
    	\includegraphics[width=0.7\textwidth]{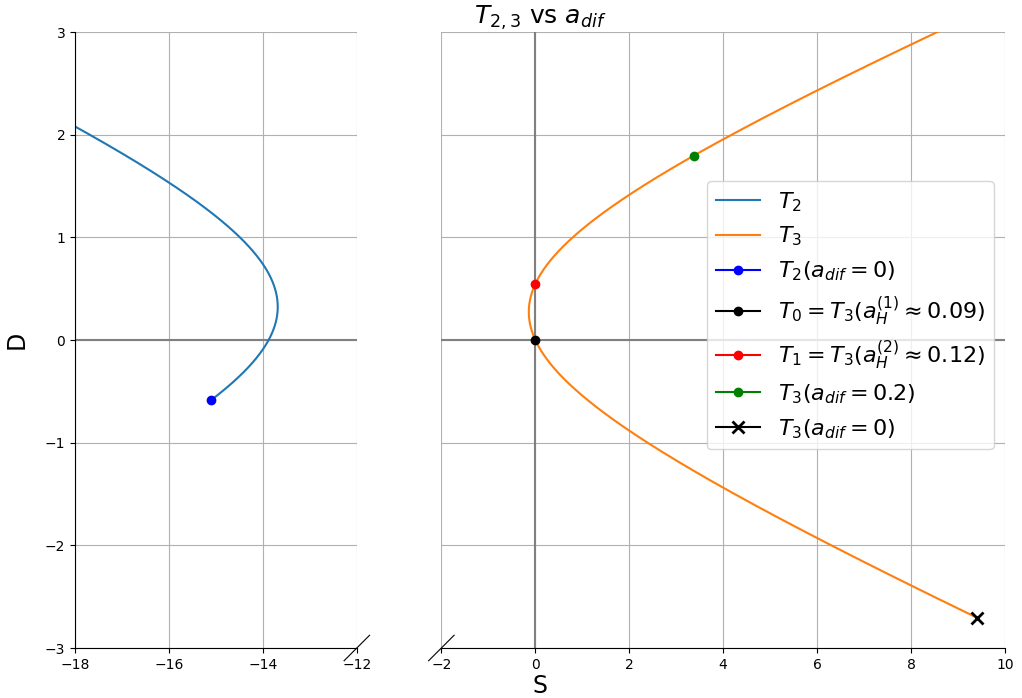}
    	\captionof{figure}{\emph{Position of the coexistence states in terms of \(a_{dif}\).} The phase space is plotted with the origin as the black dot, \(T_1\) as the red one, and the blue and orange curves as the positions of \(T_2(a_{dif})\) and \(T_3(a_{dif})\), respectively. \(T_2\) is always non-biological and \(T_3\) becomes biologically relevant for efficiencies greater than \(a^{(2)}_{H}\). The green dot corresponds to the position of \(T_3\) for \(a_{dif}=0.2\). This will be used as a standard later.} \label{figure: trayectorias T23 vs adif}	
        \end{figure}

        \subsection{Experiments of Wang et al.}  \label{Subsection Wang}

        In these experiments \cite{Wang2016}, tumorspheres were cultured on agar substrates of different stiffnesses. Here we present the results for a hard substrate, but the case of growth on a soft agar is qualitatively similar. Wang et al. obtained cultures enriched in CSCs by adding an epidermal growth factor (EGF) and a basic fibroblast growth factor (b-FGF) \cite{Wang2016}. They reported the time evolution of tumorsphere sizes, and these data were used in \cite{Benitez2021} to obtain the model parameters and analyze growth in the absence of therapy. Again, we take these parameters (see Table \ref{tab: param}) and feed them to our model in order to obtain the system dynamics as a function of \(a_{dif}\).

        In the absence of therapy \(T_0\) is a repulsor and \(T_1\) a saddle node. The evolution of the equilibria as \(a_{dif}\) is increased is shown in Fig. \ref{figure: trayectorias T23 vs adif}. \(T_2\) is always non biological. \(T_3\) starts in the fourth quadrant, crosses into the second quadrant through the origin, generating a transcritical bifurcation when it meets \(T_0\) (at \(a_{dif}=a^{(1)}_{H}\)), which then becomes a saddle node. When \(a_{dif}\) reaches \(a^{(2)}_{H}\), \(T_3\) enters the first quadrant generating a new transcritical bifurcation as it crosses \(T_1\). As a result, for \(a_{dif}>a^{(2)}_{H}\), \(T_3\) becomes a saddle node and \(T_1\) an attractor. We summarize the biologically relevant behaviors in Table \ref{tab: regimenes hard}.

        \begin{table*}
    	\centering
    	\begin{tabular}{ |p{1.5cm}||p{2cm}|p{2.5cm}|p{2.5cm}|p{2.5cm}|p{2.5cm}|  }
    		\hline
    		Case& Range & \(T_0\) &\(T_1\)&\(T_2\)&\(T_3\)\\
    		\hline
    		1-A & \([0,a^{(1)}_{H})\)  & Repulsor & Saddle Node & Non-biological & Non-biological\\
    		1-B & \((a^{(1)}_{H},a^{(2)}_{H})\) & Saddle Node & Saddle Node & Non-biological & Non-biological\\
    		2 & \((a^{(2)}_{H},\infty)\) & Saddle Node & Attractor & Non-biological & Saddle Node\\
    		\hline
    	\end{tabular}
    	\caption{Summary of the qualitative behavior of the system, for different ranges of \(a_{dif}\), with the parameters of the experiment of Wang et al. \cite{Wang2016} on a hard substrate.}
    	\label{tab: regimenes hard}
        \end{table*}

\begin{figure}
    	\centering
    	\begin{subfigure}[b]{0.49\textwidth}
    		\centering
    		\includegraphics[width=1\textwidth]{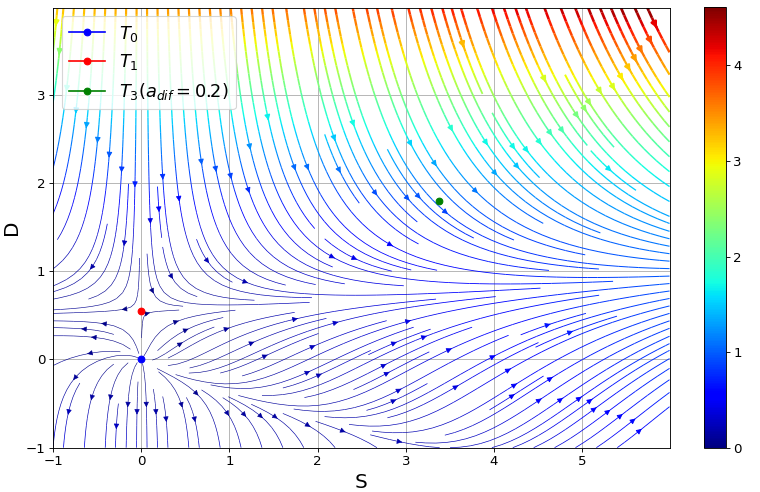}
    		\caption{Phase portrait for \(a_{dif} = 0\) (case 1-A).}
    		\label{fig: DiagFluj Hard Reg1}
    	\end{subfigure}
    	\hfill
    	\begin{subfigure}[b]{0.49\textwidth}
    		\centering
    		\includegraphics[width=1\textwidth]{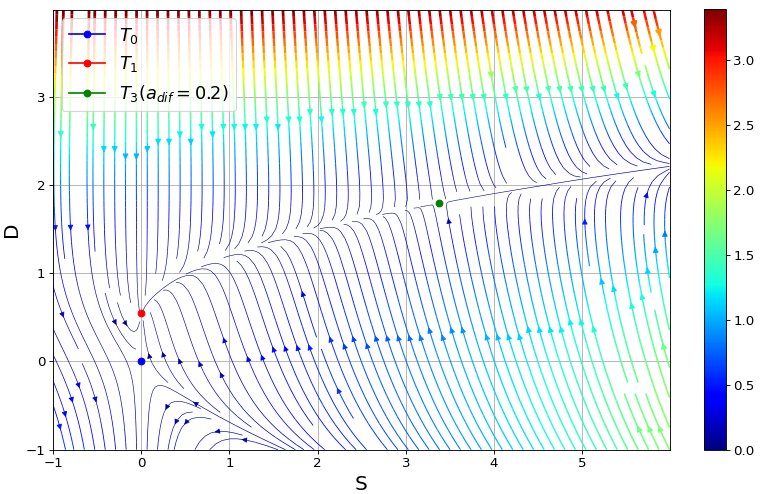}
    		\caption{Phase portrait for \(a_{dif} = 0.2\) (case 2).}
    		\label{fig: DiagFluj Hard Reg2}
    	\end{subfigure}
    	\caption{\textit{Phase portraits, hard substrate (Wang).} A typical phase portrait is displayed for each case. The lines follow a color code that indicates the velocity of the flux (given by the module of the derivative vector, in units of 1/days—see sidebar). The origin \(T_0\) and the \(D\)-cell equilibrium \(T_1\) are also plotted. \(T_3(a_{dif}=0.2)\) is also indicated in panel \ref{fig: DiagFluj Hard Reg1} for comparison purposes.}
    	\label{fig: DiagFluj Hard}
        \end{figure}

\begin{figure*}
    	\centering
    	\begin{subfigure}[b]{0.49\textwidth}
    		\centering
    		\includegraphics[width=1\textwidth]{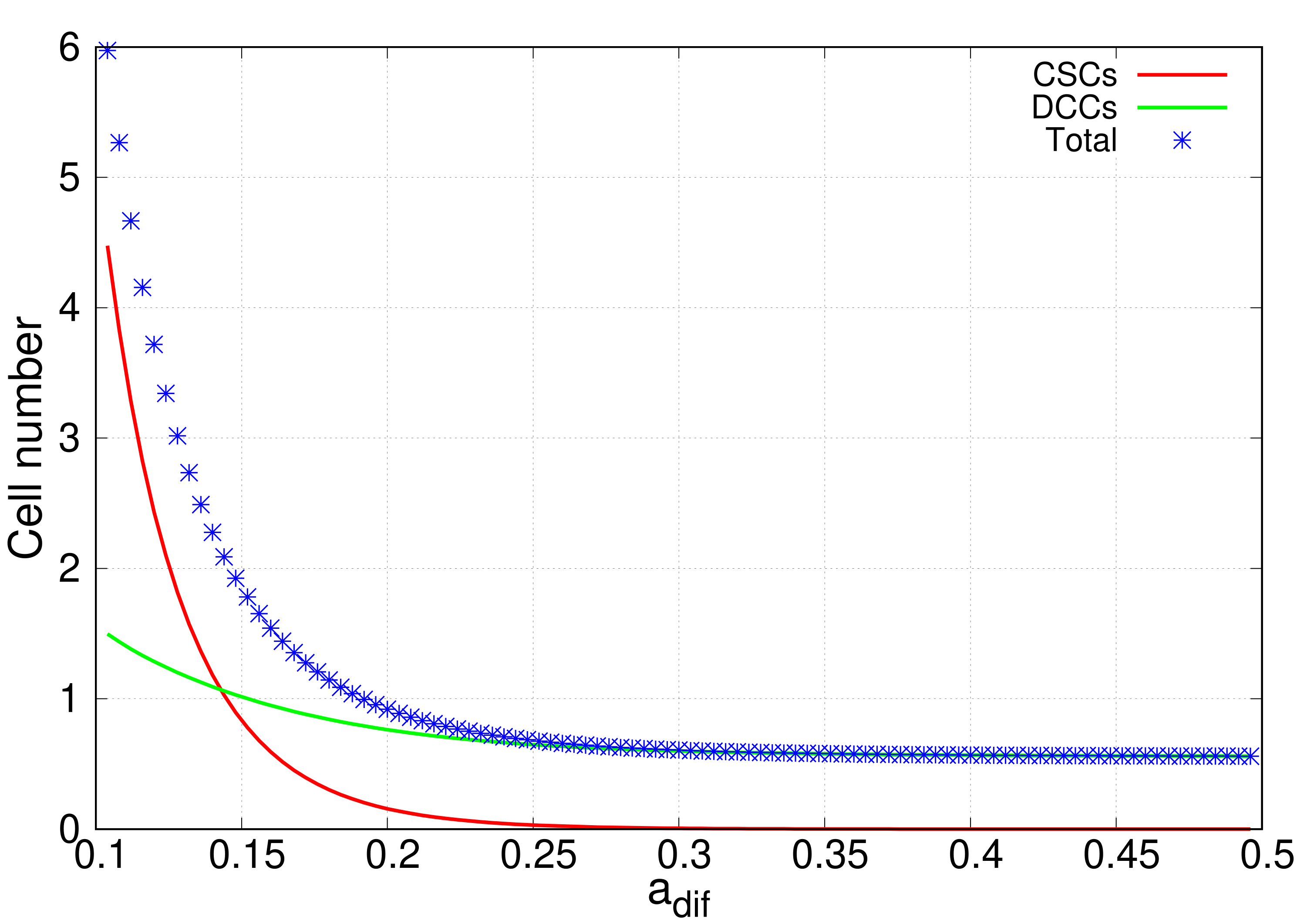}
    		\caption{Populations at \(t = 30\) days vs \(a_{dif}\).}
    		\label{fig: Umbral adif tf 30}
    	\end{subfigure}
    	\hfill
    	\begin{subfigure}[b]{0.49\textwidth}
    		\centering
    		\includegraphics[width=1\textwidth]{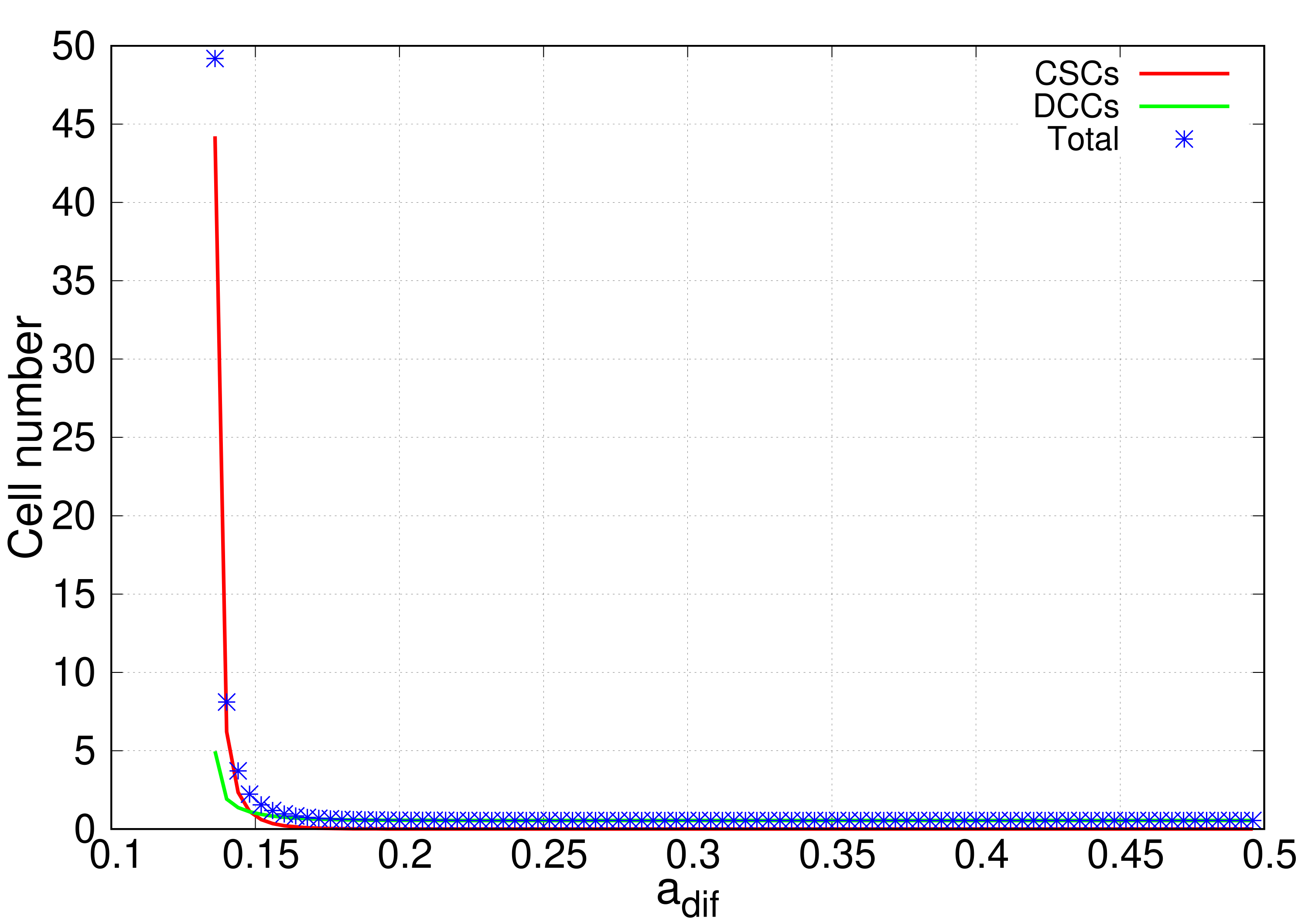}
    		\caption{Populations at \(t = 80\) days vs \(a_{dif}\).}
    		\label{fig: Umbral adif tf 80}
    	\end{subfigure}
    	\hfill
    	\begin{subfigure}[b]{0.49\textwidth}
    		\centering
    		\includegraphics[width=1\textwidth]{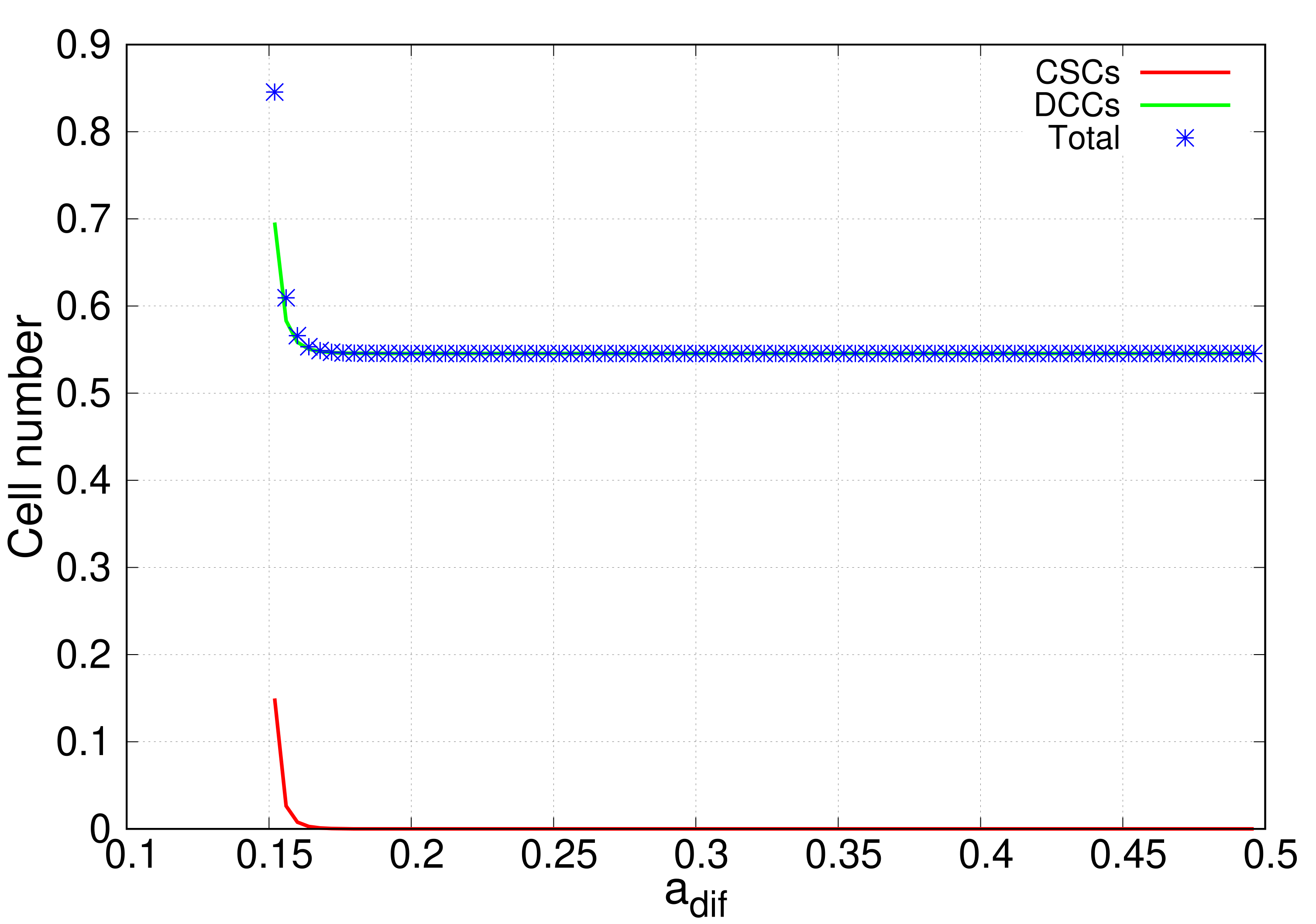}
    		\caption{Populations at \(t = 180\) days vs \(a_{dif}\).}
    		\label{fig: Umbral adif tf 180}
    	\end{subfigure}
    	\hfill
    	\begin{subfigure}[b]{0.49\textwidth}
    		\centering
    		\includegraphics[width=1\textwidth]{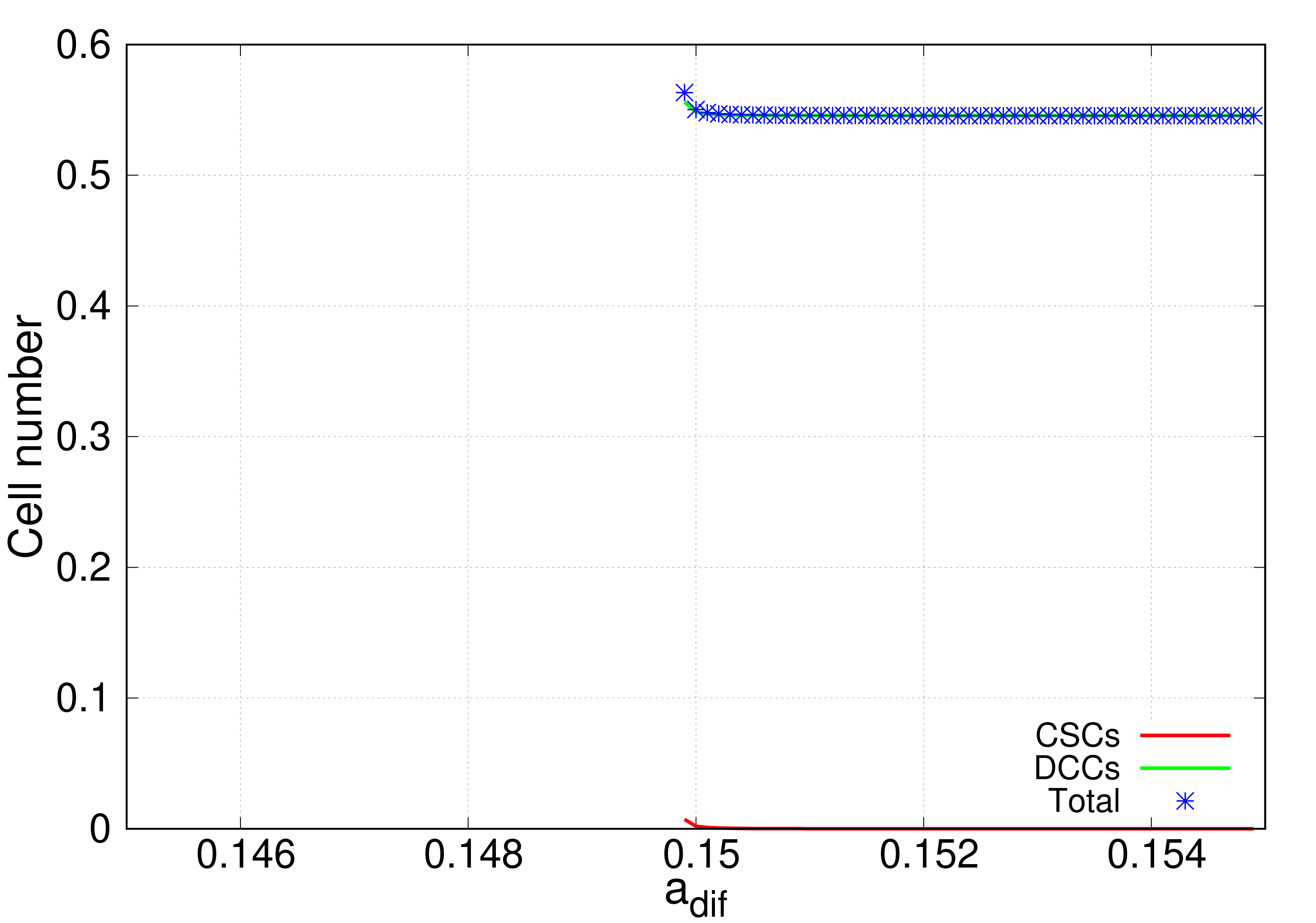}
    		\caption{Populations at \(t = 500\) days vs \(a_{dif}\).}
    		\label{fig: Umbral adif tf 500}
    	\end{subfigure}
    	\caption{\emph{Threshold formation for obtaining \(a^{min}_{dif}\).} Populations at various times \(t\) are shown as functions of \(a_{dif}\), for the initial condition \((S_0, D_0) = (1, 0)\). As \(t\) increases, the threshold becomes sharper, revealing the value of \(a^{min}_{dif}\). In panel \ref{fig: Umbral adif tf 500} the value of \(a^{min}_{dif}\) is seen to be slightly lower than \(0.15\). Populations for lower therapy efficiencies are very large and fall out of range.}
    	\label{fig: Umbral adif}
        \end{figure*}

        We observe that there are again two cases. In case 1, every initial condition in the first quadrant leads to divergent solutions, that is, solutions whose growth is unbounded. In case 2, the first quadrant is divided by the stable manifold of \(T_3\) in the basin of attraction of \(T_1\), and the domain outside the basin, where any initial condition leads to unbounded growth. Both cases are illustrated with examples in Fig. \ref{fig: DiagFluj Hard}, where we see that, by increasing \(a_{dif}\), \(T_3\) is shifted farther away from the origin. This increases the size of the basin of attraction of \(T_1\), as its boundary is the stable manifold of \(T_3\). As a result, \(a_{dif}\) controls which initial conditions lead to \(T_1\) and which lead to unlimited growth. More specifically, we can say that, given an initial condition, there is a threshold value \(a^{min}_{dif}\) such that stronger therapies force the system to converge to \(T_1\), while weaker therapies cannot restrict unbounded growth. This threshold value corresponds to the efficiency for which the initial condition belongs to the stable manifold of \(T_3\). One way to obtain its value is to plot the population at late times as a function of \(a_{dif}\), for a given initial condition. This can be seen in Fig. \ref{fig: Umbral adif}, where we can observe how the threshold becomes more definite as the observation time \(t\) increases.

\begin{figure}
    	\centering
    	\includegraphics[width=0.6\textwidth]{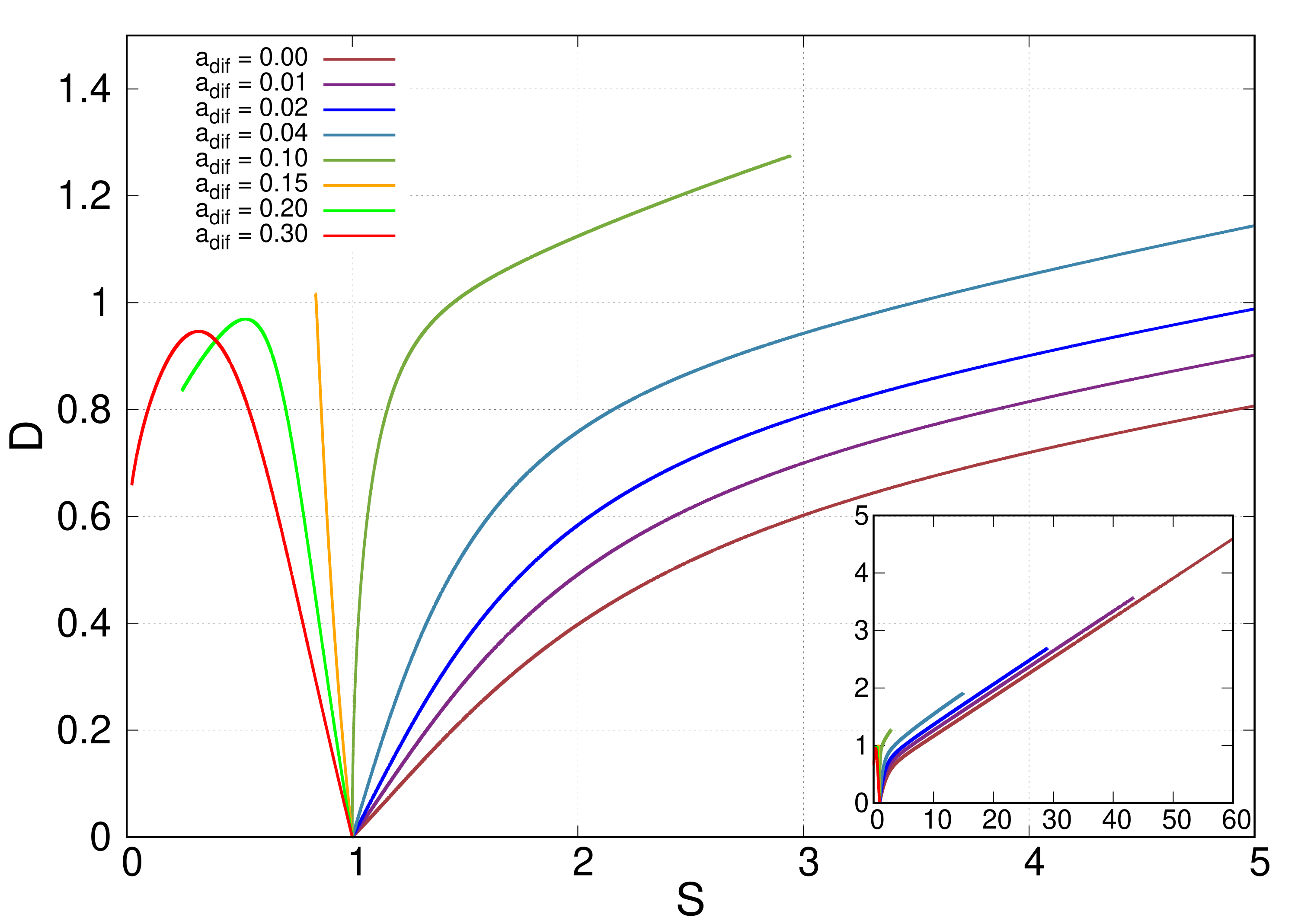}
    	\captionof{figure}{\emph{Example illustrating the threshold \(a^{min}_{dif}\) for a single CSC as seed.} Trajectories in the phase space for the first 23 days of evolution from a single CSC, for different therapy strengths. For strengths higher than \(a^{min}_{dif} \approx 0.15\) the system converges to \(T_1\), while for lower strenghts it grows without limits. Growth is faster for smaller \(a_{dif}\).} \label{figure: efecto adif hard misma CI}	
        \end{figure}

         In Fig. \ref{figure: efecto adif hard misma CI} we show  the first 23 days of the system's evolution out of the initial condition \((S_0, D_0) = (1, 0)\), for different values of \(a_{dif}\). For this initial condition, \(a^{min}_{dif}\) is slightly lower than \(0.15\) (as shown in Fig. \ref{fig: Umbral adif}). Therefore, the orange curve corresponding to \(a_{dif}=0.15\) eventually gets to \(T_1\) and serves as a reference for the change in behavior. Curves with \(a_{dif}>a^{min}_{dif}\) end in \(T_1\), while the ones with \(a_{dif}< a^{min}_{dif}\) diverge.

        If we used Theorem \ref{Teo: extension teo 2} to estimate the threshold, we would find that it gives a poor approximation, \(r (p_s-p_d) = 0.095\). The reason for this is that, as anticipated in the remarks, a single CSC is not a small seed in this case, due to the large value of \(\alpha_{DD}\).

        It is worth noting that, even when growth is unbounded, therapy manages to slow it down. This can be verified by noting that, at a given time, the total population increases as \(a_{dif}\) decreases (for an example, see Fig. \ref{figure: efecto adif hard misma CI}). This is reasonable, since DCCs compete with each other (\(\alpha_{DD}>0\)), while CSCs cooperate (\(\alpha_{SS}<0\)) and promote DCCs more strongly than they are promoted by them (\(\alpha_{SD}<\alpha_{DS}<0\)).

        The assumption that therapy and growth start simultaneously is quite artificial. It would be more reasonable to consider that growth starts at \(t_0=0\) and that therapy is implemented at a later time \(t_T > 0\). If we consider that \(a_{dif}(t) = \theta(t-t_T)\tilde{a}\), where \(\theta(t)\) is the Heaviside step function, we have two possibilities. For \(\tilde{a}<a^{(2)}_{H}\), the evolution of the system takes place in case 1, both before and after the beginning of the therapy. This implies that solutions diverge. If \(\tilde{a}>a^{(5)}_{H}\), there is a transition from case 1 to case 2 after therapy starts. The growth of the system will be unbounded only if, by time \(t_T\), it manages to escape the basin of attraction of \(T_1\) that corresponds to \(\tilde{a}\).

        Again, given an initial condition and a sufficiently large value of \(\tilde{a}\), there will be a threshold value \(t^{Max}_T\) such that \(t_T > t^{Max}_T\) implies that growth in unbounded, while \(t_T < t^{Max}_T\) implies that the system converges to \(T_1\). This value is given by the time at which the system gets to the stable manifold of \(T_3(\tilde{a})\), evolving in the absence of therapy from a given initial condition. Note that if \(\tilde{a}\) is not large enough, the initial condition may not be in the basin of attraction of \(T_3(\tilde{a})\), and therefore growth is unbounded regardless of the therapy starting time.

        \begin{figure}[h]
		\centering
		\begin{subfigure}[b]{0.49\textwidth}
			\centering
			\includegraphics[width=1\textwidth]{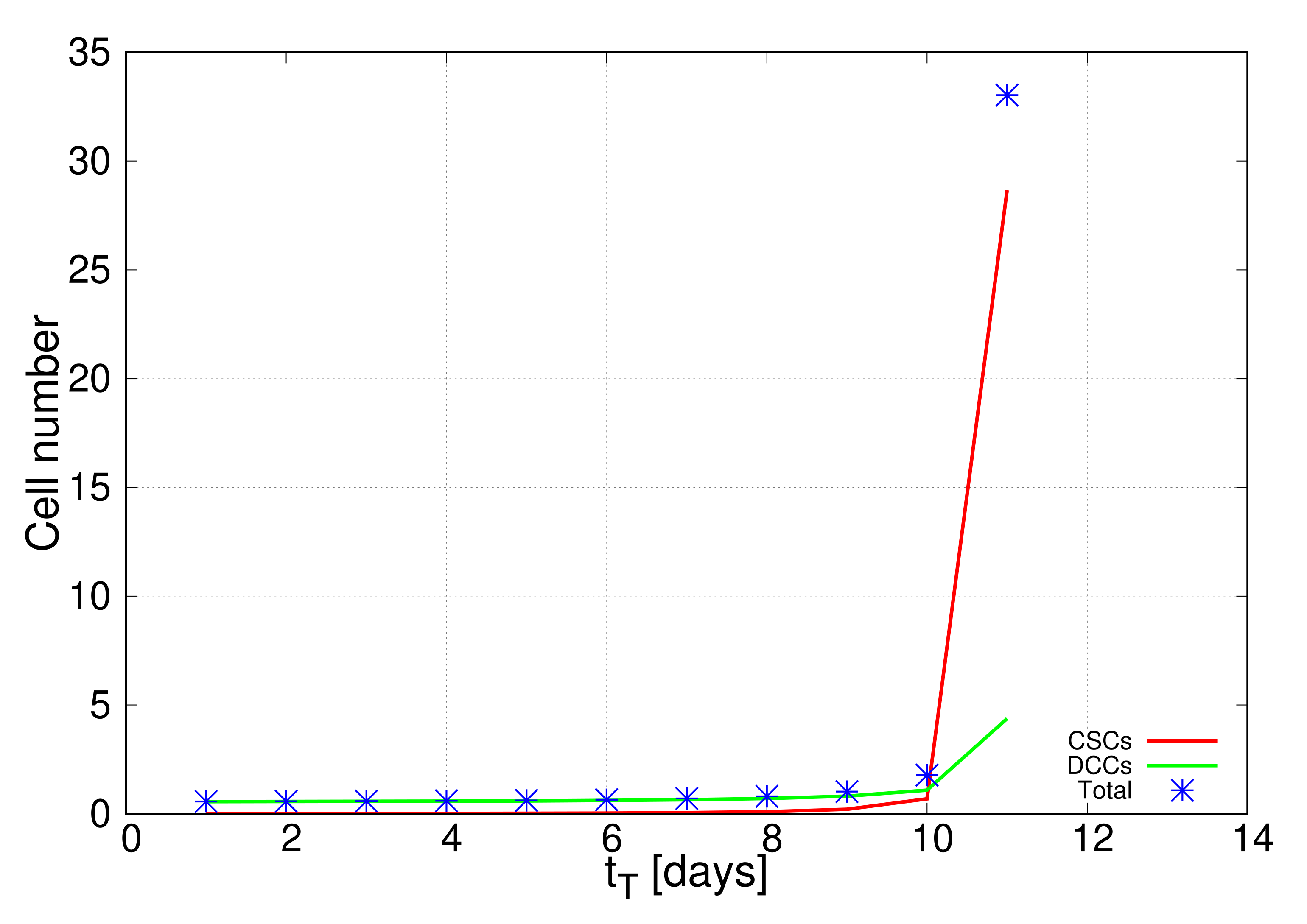}
			\caption{Populations at \(t = 80\) days vs \(t_{T}\).}
			\label{fig: Umbral tt tf 80}
		\end{subfigure}
		\hfill
		\begin{subfigure}[b]{0.49\textwidth}
			\centering
			\includegraphics[width=1\textwidth]{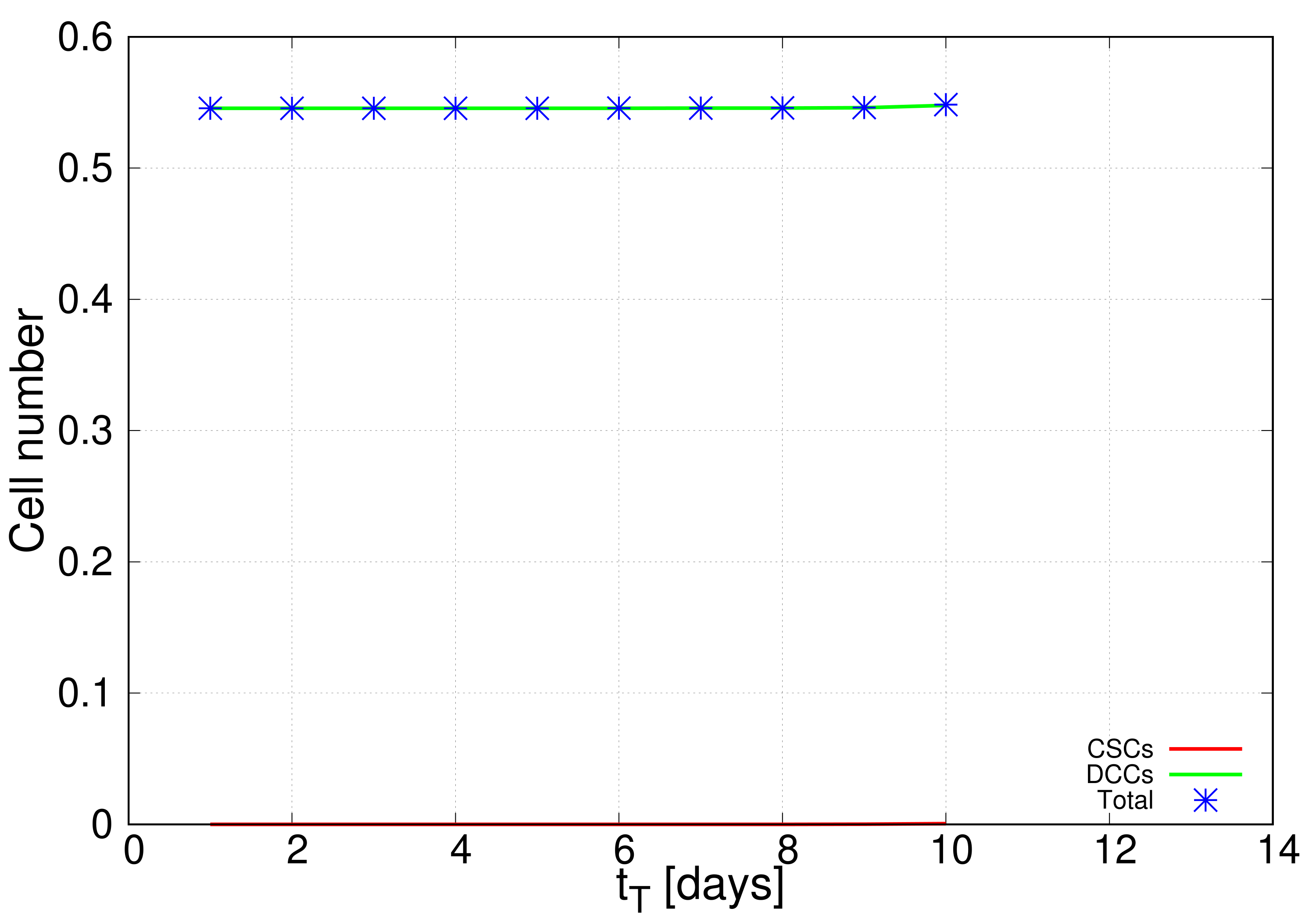}
			\caption{Populations at \(t = 180\) days vs \(t_{T}\).}
			\label{fig: Umbral tt tf 180}
		\end{subfigure}
		\caption{\emph{Step formation for obtaining the threshold value \(a^{min}_{dif}\).} Numbers of CSCs (red), DCCs (green) and total number of cells (blue) as functions of \(t_T\) for (a) \(t=80\) days and (b) \(t=180\) days, with \(\tilde{a}=0.2\). Results were calculated for integer values of \(t_T\), the solid lines being a guide to the eye. The initial condition is \((S_0, D_0) = (1, 0)\). As \(t\) increases, the step defining the threshold \(t^{Max}_T\) becomes more sharply defined: for \(t_T < t^{Max}_T \approx 10\) days the subpopulations converge to \(T_1 \approx (0, 0.5)\), while for \(t_T > t^{Max}_T \approx 10\) days they are beyond the edges of the figure.}
		\label{fig: Umbral tt}
	\end{figure}

        \begin{figure}[h]
		\centering 
		\begin{subfigure}[b]{0.49\textwidth}
			\centering
			\includegraphics[width=1\textwidth]{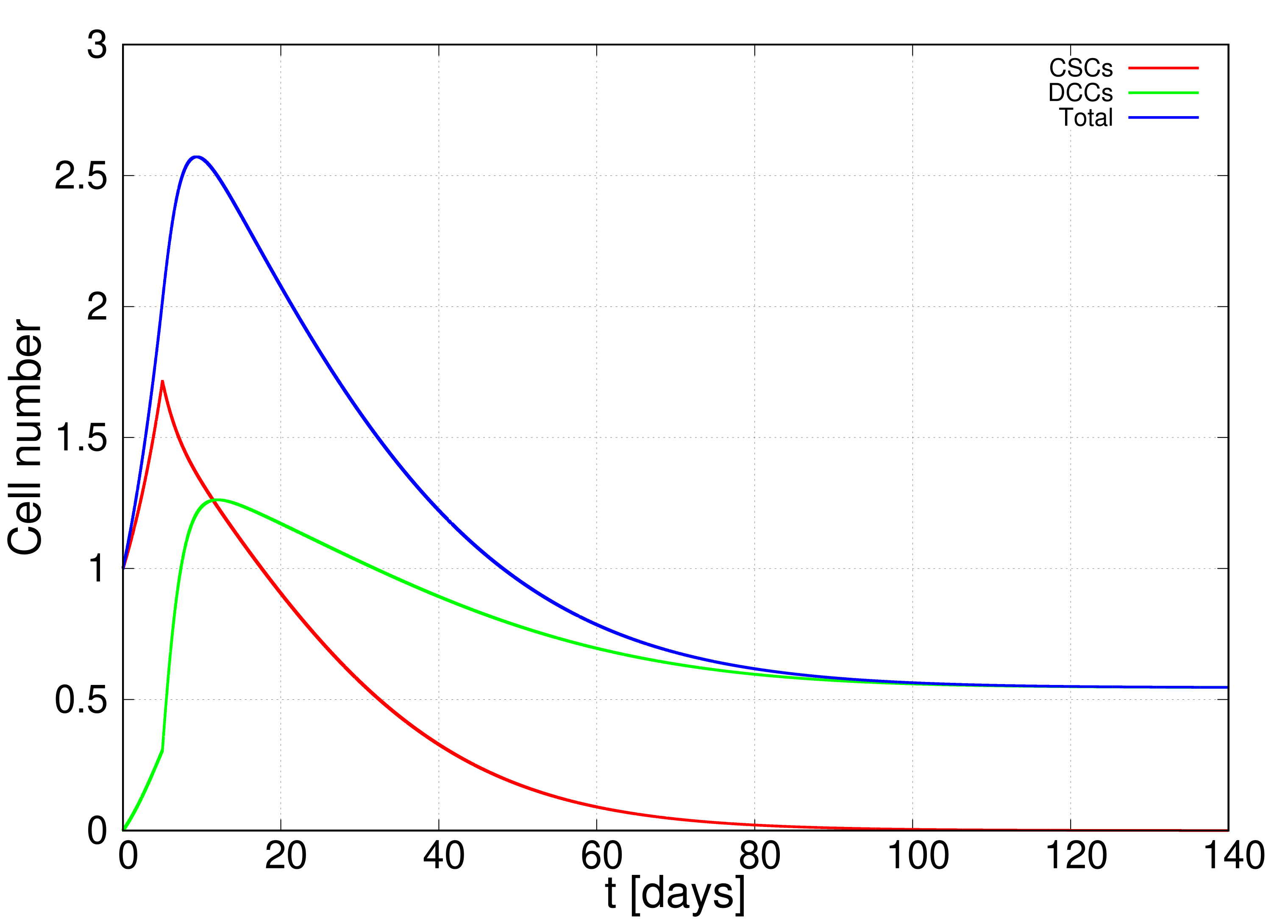}
			\caption{Evolution with \(t_{T} = 5\) days.}
			\label{fig: Umbral tt 5}
		\end{subfigure}
		\hfill
		\begin{subfigure}[b]{0.49\textwidth}
			\centering
			\includegraphics[width=1\textwidth]{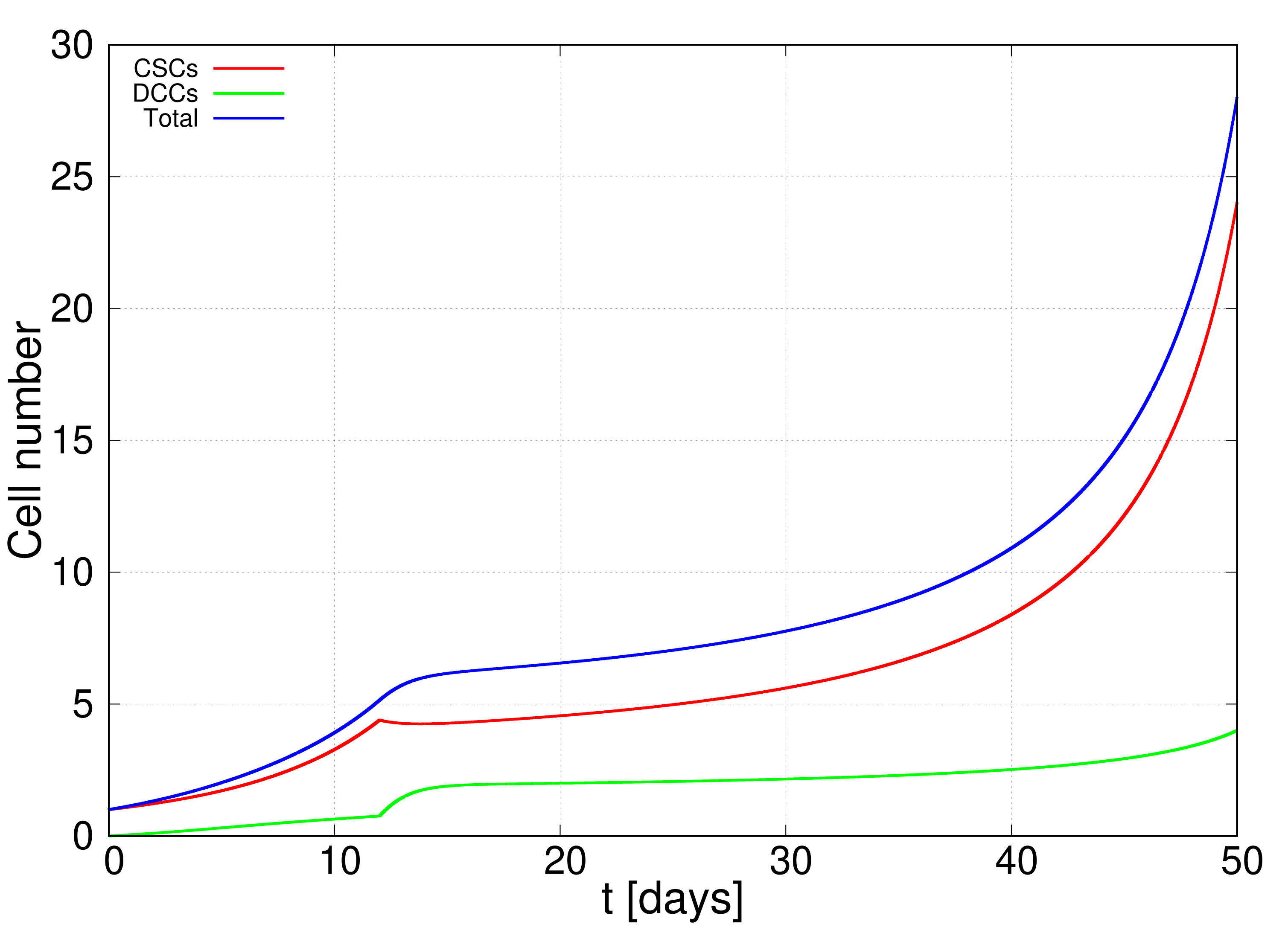}
			\caption{Evolution with \(t_{T} = 12\) days.}
			\label{fig: Umbral tt 12}
		\end{subfigure}
		\caption{\emph{Early vs a late therapy starts.} Time dependence of the tumorsphere subpopulations for the initial condition \((S_0, D_0) = (1, 0)\) and \(\tilde{a}=0.2\). Therapy starts at (a) \(t_{T} = 5\) days and (b) \(t_{T} = 12\) days. In (a) the therapy onset occurs before the threshold and leads to a CSC-free sphere, while in (b) the therapy begins after the threshold and can only slow down growth.}
		\label{fig: Umbral tt ejemplo}
	\end{figure}

        As an example, let us analyze the case \((S_0, D_0) = (1, 0)\) and \(\tilde{a}=0.2\). To find the threshold, we plot the populations at time \(t\) as functions of \(t_T\) (see Fig. \ref{fig: Umbral tt}). For large \(t\), \(D(t_T)\) tends to a step function with its non-differentiable point at the threshold \(t^{Max}_T\); if \(t_T < t^{Max}_T\) the system tends to \(T_1\); otherwise, it grows without limits. This change is rather sharp: In Fig. \ref{fig: Umbral tt ejemplo} we show both subpopulations as functions of time for \(t_T = 5 \text{ days} < t^{Max}_T \simeq 10 \text{ days}\), i.e. an early beginning of the therapy, and for \(t_T = 12 \text{ days} > t^{Max}_T \approx 10 \text{ days}\), a late beginning. The cusp in both subpopulations at the time of the therapy onset disappears for the total population (the therapy only transfers members from one subpopulation to the other).

\begin{figure}
    	\centering
    	\includegraphics[width=0.5\textwidth]{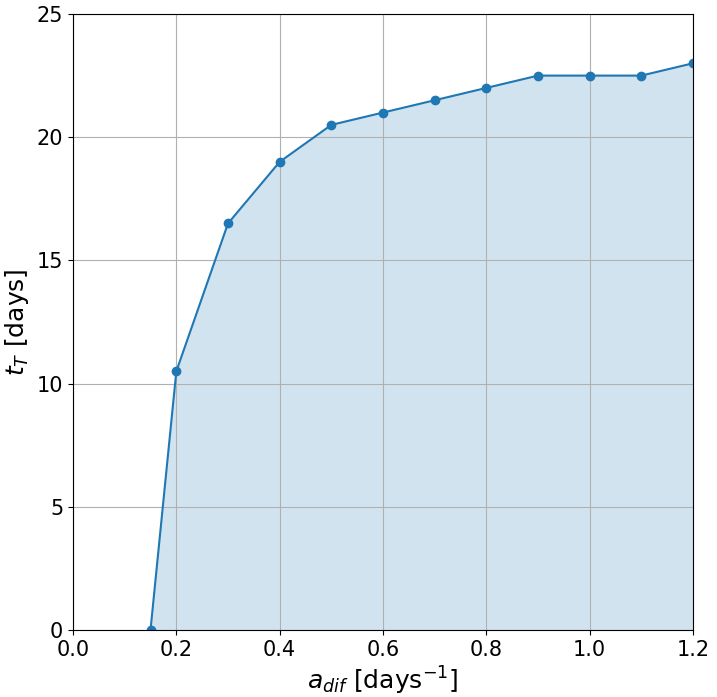}
    	\caption{\emph{Latest starting time as a function of therapy strength for the parameters of the experiment on a hard substrate.} The curve \(t_T^{Max}(a_{dif})\) is plotted. Points \((a_{dif}, t_T)\) below this curve correspond to solutions that converge to \(T_1\), while points above the curve give divergent solutions (the therapy fails).}
    	\label{figure: curva ttmax vs adif HARD}
        \end{figure}

        By finding \(t^{Max}_T\) for different efficiencies, we can now draw a curve \(t_T^{Max}(a_{dif})\) that separates the therapy efficiencies and starting times that lead to divergent solutions, from those that make the system converge to \(T_1\). Using a single CSC as the initial condition, the resulting curve is shown in Fig. \ref{figure: curva ttmax vs adif HARD}, where the parameter space is divided in combinations that lead to divergent solutions and combinations that lead to solutions that tend to \(T_1\). It is worth noting that, by increasing the therapy strength, we can delay its beginning and still control growth. Besides, the intersection of the curve with the horizontal axis takes place at \(a^{min}_{dif}\), showing that there is an absolute minimum efficiency needed for any given initial condition. This value will always be greater than that necessary to be in case 2, namely \(a^{(2)}_{H}\).

        Comparison with the case discussed in \ref{Subsection Chen} suggests that the effect of the additives EFG and b-FGF is to remove \(T_2\) from the first quadrant, i.e., they prevent the emergence of coexistence states. In the case depicted in Fig. \ref{fig: Umbral tt}, we see that the \(D\)-cell equilibrium \(T_1\) contains less than one cell, which indicates that DCCs cannot survive without the assistance provided by the CSCs. In this case, the differentiating agent is likely to eliminate the tumor completely.

        The difference between the two experimental cases exposed may be explained by the availability of a substrate on which to grow. Free growth is always bounded, while growth on a substrate can be unbounded. Substrate stiffness can accelerate or delay growth, but produces no qualitative difference in the behavior. This is supported by the fact that using the parameters for the \textit{``soft''} substrate (results not shown) produces results qualitatively identical to those obtained using the parameters for the \textit{``hard''} substrate. In any case, culture conditions are likely to determine whether there will be uncontrolled growth.

        \section{Conclusions}

        We generalized the model of Benítez et al. to investigate the effects of the application of a differentiating agent to a tumorsphere. We studied the properties of the extended model, found the equilibria, and proved two theorems regarding the positivity of the solutions, the conditions for initial growth, and some properties of the solutions.

        We then used the model to make predictions about the effects of therapy on tumorspheres grown under very different experimental conditions. First, for spheres grown in a microchamber \cite{Chen2016}, we found that for low therapy strengths the tumorsphere ends up consisting of a mixture of the two subpopulations, while for high strengths the CSC fraction disappears. The effect of the therapy is to reduce tumorsphere size, its starting time not being relevant. Second, for tumorspheres cultured on substrates of different stiffnesses \cite{Wang2016}, we found that, if growth is to be controlled, there is a minimum therapy strength for any initial condition; in the case of a hard substrate, this strength corresponds to the bifurcation value \(a^{(2)}_H\). In the absence of therapy, the solutions are unbounded due to the large interspecific cooperation; control can only be achieved by eliminating the CSC fraction. We also found the latest starting time for a successful therapy of a given strength, or equivalently, the minimum strength to control growth given a starting time.

        The differences between the two cases underscore the importance of the environment, suggesting that the presence of a substrate may be responsible for uncontrolled growth. The model shows then how a differentiation therapy outcome would not only depend on dosage and timing, but also, critically, on the tumor microenvironment. The work presented here should be extended to incorporate \textit{in vivo} conditions and the effects of conventional therapies, but our results strongly suggest that the effect of the differentiating therapy on a real tumor would be better approximated by those in Section \ref{Subsection Wang}, since substrate adhesion and stemness-promoting agents are likely to be present.

        \section*{Acknowledgments}
        This work was supported by SECyT-UNC (Project 113/17) and CONICET (PIP 11220110100794), Argentina. We thank Lucía Benítez and Luciano Vellón for useful discussions.


\bibliographystyle{elsarticle-num} 
\bibliography{bibliography}


\end{document}